Ananya Bhattacharjee, PhD[1,2]
Jina Suh, PhD[3]
Mohit Chandra, MS[4]
Javier Hernandez, PhD[3]

[1]King Center on Global Development, Stanford University, Stanford, CA, USA
[2]Department of Computer Science, University of Toronto, Toronto, ON, Canada
[3]Microsoft Research, Redmond, WA, USA
[4]School of Interactive Computing, Georgia Institute of Technology, Atlanta, GA, USA

# User Perceptions of an LLM-Based Chatbot for Cognitive Reappraisal of Stress: Feasibility Study

## Abstract

**Background:** Cognitive reappraisal is a widely studied emotion regulation strategy that helps individuals reinterpret stressful situations in ways that reduce their emotional impact. Digital mental health (DMH) tools often struggle to support this process because scripted templates fail to adapt to the varied and incomplete ways users describe their stressors. Large language models (LLMs) offer opportunities to increase conversational flexibility while preserving structured intervention steps.

**Objective:** This study examined the feasibility of an LLM-based single-session intervention (SSI) for stress reappraisal. We aimed to assess whether the activity would be associated with short-term improvements in stress-related outcomes, and what design tensions arise during user interaction.

**Methods:** We conducted a feasibility study with 100 employees from a large US technology company. Participants completed a structured cognitive reappraisal session delivered by a GPT-4o–based chatbot within Qualtrics. Pre–post measures included situational stress intensity, stress mindset, perceived demand, and perceived resources (all 5-point scales). Paired Wilcoxon signed-rank tests were used with Benjamini-Hochberg correction. To complement self-reports, we analyzed sentiment and stress trajectories across conversation tertiles using a RoBERTa sentiment classifier, a RoBERTa stress classifier, and an LLM-based stress rater. We also conducted a thematic analysis of participants' open-ended responses covering stress impact, perspective change, liked and disliked aspects, and improvement suggestions, alongside their conversation logs.

**Results:** Significant reductions were observed in situational stress intensity ($\Delta$ = 0.29 ± 0.83, $p = 0.002$, $r_{rb} = 0.54$) and significant improvements in stress mindset ($\Delta$ = 1.70 ± 4.37, $p = 0.002$, $r_{rb} = 0.44$). Sentiment and stress classifiers showed consistent declines in negative sentiment and stress from conversation start to end (all omnibus Friedman tests $p < 0.001$). Thematic analysis of open-ended responses surfaced several patterns in how participants engaged with the activity: they valued the structured prompts for organizing their thoughts, described shifts in how they interpreted their stressors, used the interaction to identify concrete next steps, and experienced the chatbot as a non-judgmental space for reflection. Reported design tensions included perceived scriptedness, variable preferences for conversation length, and mixed reactions to AI-driven empathy.

**Conclusions:** An LLM-enhanced cognitive reappraisal activity showed promise to be delivered as a brief digital intervention and is associated with short-term improvements in situational stress and stress mindset. Participants appreciated the clarity and reflection supported by the structured sequence, while noting important design challenges in balancing structure with conversational naturalness and contextual depth. These findings highlight both the promise and the design constraints of integrating LLMs into DMH interventions.

## Introduction

Cognitive reappraisal is an emotion regulation strategy that involves reframing the meaning of a situation in order to change its emotional impact [1–3]. In the context of stress, this entails altering one's interpretation of a stressful event to reduce its harmful emotional effects. It is considered a form of cognitive appraisal that enables individuals to reinterpret an event, often finding new meaning or viewing the outcome as more benign [1]. By changing the way one thinks about a stressor, cognitive reappraisal can shift the emotional response from anxiety or anger to a more neutral or even positive state. For example, interpreting a professional setback as a growth opportunity rather than a failure can diminish distress and foster adaptive coping [4].

A substantial body of research demonstrates that cognitive reappraisal is a beneficial strategy for psychological and emotional well-being. Individuals who regularly use reappraisal tend to report lower levels of depression and anxiety [2,3,5]. Over time, greater reappraisal ability has also been linked to a reduced risk of developing depressive symptoms, particularly under high stress [1]. Reappraisal has been associated with more frequent positive affect and greater life satisfaction [2]. In addition, people who use reappraisal often report better interpersonal relationships [6,7]. These benefits have been observed across diverse populations and contexts, with studies showing that habitual use of reappraisal correlates with improved psychological health over time [3]. Together, these findings highlight the substantial and wide-ranging benefits of cognitive reappraisal for well-being.

However, existing strategies for cognitive reappraisal of stress, similar to many digital mental health (DMH) interventions, often struggle to provide support that feels applicable to users' situations [8]. An important challenge lies in their reliance on scripted responses and fixed templates that proceed through a predetermined sequence without adjusting to the way users naturally describe their experiences [8,9]. Individuals may give partial descriptions, mention several elements of the situation at once, or provide information that is either too broad or too narrow for the next required step in the intervention. Scripted tools (e.g., [10–12]) typically move forward regardless of these variations, which can create moments where the guidance feels rigid or misaligned with what the user is attempting to communicate. Prior chatbots such as Woebot [13] and Wysa [14] demonstrated that structured CBT-style exchanges could produce measurable reductions in stress and negative affect, though their scripted templates limited contextual responsiveness. These limitations have highlighted the need for systems that maintain a consistent intervention structure while allowing conversational adjustments within each step. Approaches that combine structured reappraisal steps with conversational responsiveness can help reduce the sense of mismatch that users often report with scripted digital interventions [15,16].

Large language models (LLMs) offer an emerging opportunity to address limitations of traditional stress reappraisal tools that rely on fixed scripts and rigid templates [17,18]. Instead of producing predetermined responses, LLMs can engage in brief conversational exchanges that acknowledge users' descriptions and request clarification when needed [19,20]. This capacity supports a smoother dialogue within a structured intervention, helping the system remain connected to what the user has already expressed. For example, if a user introduces a stressor in broad terms, an LLM-based agent can provide a simple acknowledgment and ask for a small detail required for the next step. If a user highlights a particular source of tension, the agent can recognize that element before guiding the user forward in the predefined sequence. These conversational elements can help maintain coherence between the user's wording and the transitions across steps, which may reduce the sense of rigidity that often occurs in scripted tools [17,21].

While LLMs offer exciting possibilities, they cannot simply be inserted into well-being interventions such as cognitive reappraisal without careful design. These interventions are inherently structured, and preserving that structure is essential for delivering appropriate support. Without it, there is a risk of generating content that may be unhelpful or even counterproductive. Rather than viewing LLMs as replacements for structured interventions, we posit that their promise lies in augmenting these techniques—enabling context-sensitive interaction through natural language. By acknowledging the user's situation and adapting responses accordingly, LLMs may help structured interventions feel tailored, while still upholding the core principles that guide them [17,19,22].

Natural Language Processing (NLP) research has used language models to support reframing of negative thoughts, including positive reframing as style transfer [23] and identification and reframing of cognitive distortions [21]. More recent work has directly examined LLM-based cognitive reappraisal: Sharma et al. [22] showed that LLMs can support identification of thinking traps and provision of contextually appropriate reframes, and Zhan et al. [16] showed that LLMs guided by psychologically grounded constitutions can generate targeted reappraisals rated by clinical psychologists as more aligned with established principles than human-written responses. However, these approaches have largely focused on generating individual reappraisal responses, with less attention to the multi-turn conversational process through which reframing often unfolds in real-world supportive interactions.

Motivated by these opportunities and challenges, we examined the feasibility of an LLM-enhanced cognitive reappraisal intervention delivered as a single session intervention (SSI). SSIs are brief, structured activities that can produce rapid shifts in thoughts or feelings and have shown benefits for stress, negative thinking, and related outcomes across varied populations [8,22,24,25]. They are well-suited for early stage exploration because they allow researchers to assess whether a short, structured activity can support users as they work through a stressful scenario. In this study, we focus specifically on situational stress, the immediate psychological tension a person experiences when confronted with a specific stressor, such as a difficult interaction with a colleague, an unresolved conflict, or an overwhelming workload. This is distinct from general perceived stress, which reflects a person's cumulative sense of feeling overwhelmed or unable to cope across their daily life over a period of weeks or months [26].

Our work was guided by the following research questions:

- **RQ1**: How does engaging with an LLM-based chatbot for cognitive reappraisal of a stressful scenario impact individuals' situational stress levels?
- **RQ2**: What design tensions arise when individuals interact with an LLM-based chatbot to cognitively reappraise a stressful scenario?

To examine these questions, we designed an SSI that used an LLM to guide users through a structured cognitive reappraisal activity [24,25]. The LLM was instructed to follow a predefined sequence similar to prior work [8]. Within each step, it incorporated conversational adaptivity by acknowledging user input and requesting brief clarifications when needed, while preserving the structure of the intervention. We chose a feasibility study design to examine whether this approach, in which the LLM facilitates a structured reappraisal process instead of generating reappraisals on the user's behalf, could be delivered reliably and produce short-term shifts in situational stress. Our aim was to gather early evidence of perceived benefits and preliminary impact, and to identify design tensions that emerge when structured psychological interventions are delivered through conversational interfaces. Accordingly, our focus was on understanding user experiences and implementation considerations rather than conducting a controlled evaluation. Feasibility work is important when introducing new technological mechanisms because it can reveal practical factors that influence real-world deployment and highlight design considerations [27,28].

A feasibility study with 100 employees from a large technology company showed that participants found the activity helpful for working through stressful situations. Qualitative feedback indicated that the structured sequence supported clearer thinking by helping participants slow down, articulate their thoughts, and reinterpret their stressors in a more constructive way. Many also described feeling validated, noting that the interaction helped them organize their emotions and provided a non-judgmental space to reflect. These qualitative insights were complemented by promising quantitative results, with significant reductions in immediate stress ($p = 0.002$, $r_{rb} = 0.54$) and improvement in stress mindset ($p = 0.002$, $r_{rb} = 0.44$), further corroborated by sentiment analysis of conversation data. However, the integration of LLMs also introduced design tensions: contextualizing the experience required additional user input, and following a predetermined structure could reduce the perceived conversational quality of the interaction.

Our contributions include (a) a structured LLM-enhanced cognitive reappraisal intervention for managing stress, (b) empirical insights into how the intervention impacts users' experiences of stress, and (c) design tensions to inform future LLM-based DMH interventions. Our study demonstrates the potential of LLM-enhanced reappraisal as a brief, scalable intervention, while also drawing attention to design challenges, including how to balance conversation length, contextual richness, and expression of empathy.

## Methods

### Design of the Intervention

#### *Theoretical Grounding and Sequence Design*

Our intervention was designed as an SSI [24,25]. SSIs are structured, standalone activities intended to produce immediate effects on targeted psychological outcomes. Typically lasting 10–20 minutes, they offer a time-efficient way to promote reflection, reappraisal, or symptom relief, making them suitable for a wide range of populations. In addition to offering rapid support, SSIs can serve as a gateway to more comprehensive mental health care by helping individuals build insight or skills that increase their willingness to seek further support [8]. Prior research has demonstrated their effectiveness in addressing negative thought patterns, reducing stress, and alleviating symptoms of anxiety and depression in both clinical and non-clinical settings [8,22,24,25].

Unlike multi-session interventions, which typically unfold over multiple weeks through repeated check-ins, skill-building exercises, and progressive assignments [12,29], SSIs are completed in a single sitting without requiring prior sessions or follow-up engagement. While multi-session programs rely on cumulative exposure and gradual skill development over time, SSIs are designed to deliver focused support within a bounded interaction. Prior work illustrates how SSIs can compress core therapeutic elements into a single experience. For example, Schleider and Weisz [30] tested a brief computer-based SSI in which participants engaged with psychoeducational content and a short reflective writing task within one sitting, and observed reductions in depression and anxiety symptoms at follow-up. More recently, Sharma et al. [20] demonstrated that cognitive reframing can be supported through structured, single-session interactions with language models. In contrast to multi-session formats that distribute these elements across time, SSIs integrate learning, reflection, and application into a single, self-contained interaction.

We drew particular inspiration from a prior SSI by [8]. The activity was designed to be easily referenced or practiced by individuals when encountering negative thoughts and emotions in their everyday lives. Their work adapted core concepts from CBT, specifically thought records and behavioral chaining [31,32], into a guided digital reflection activity. In CBT, thought records involve documenting a situation, the thoughts, emotions, and behaviors it evokes, and then generating alternative thoughts [31]; behavioral chaining focuses on tracing the sequence from trigger to reaction to identify patterns [32]. Both techniques help individuals become more aware of their internal responses and prepare for similar situations in the future by promoting insight and reflection. Given that cognitive reappraisal is not inherently context-specific, our activity was similarly designed to be broadly applicable across a range of stressful situations, with the workplace serving as the deployment context for this study.

Our activity follows a similar structure, consisting of 11 questions. The activity was developed through an iterative process involving researchers with expertise in HCI, psychology, cognitive science, generative AI, and DMH. Members of the research team had prior experience designing DMH interventions both with and without the use of AI, and have published extensively in leading HCI, psychology, and AI venues. The first set of prompts (Questions 1–7) operationalizes the cognitive model underlying CBT thought records, in which a situation gives rise to automatic thoughts that in turn drive emotional and behavioral responses [31,32]. Moving from the situation (Q1) to its most troubling element (Q2), to automatic thoughts (Q3–Q4), to feelings (Q5), to behaviors (Q6), and finally to an integrated trigger → thought → feeling → behavior summary (Q7) mirrors the standard structure used to build awareness of the links among cognition, affect, and action before any restructuring is attempted [15]. Questions 8–11 then enact cognitive reappraisal as defined in the stress-appraisal and emotion-regulation literatures [3,5]. Q8 corresponds to examining the evidence for and proportionality of an appraisal, a core component of cognitive restructuring. Q9 draws directly on the challenge-versus-threat distinction in transactional models of stress appraisal [4], inviting users to reframe a stressor as a challenge to be met rather than a threat to be avoided. Q10 reflects reappraisal in Gross's process model of emotion regulation, in which the meaning of a situation is reinterpreted to alter its emotional impact [5]. Q11 supports generalization and consolidation, encouraging users to carry the reappraisal into future situations [3]. Reappraisal comes last in the sequence by design: examining and reframing an appraisal is difficult until the cognition, emotion, and behavior tied to the trigger have been laid out, which is the work of Q1–Q7.

Table 1 shows the full sequence of prompts and their intended purpose. After completing all steps, the chatbot was instructed to provide a brief summary of the conversation and clearly state that the structured part has concluded.

Table 1: Reflection questions used in the intervention and their intended purpose

| Q# | Question | Purpose |
|---|---|---|
| 1 | What is the situation? Feel free to explain it in as much detail as you would like. | Provides context for the activity |
| 2 | What part of the situation is the most troubling? | Sets an agenda for the rest of the activity |
| 3 | What are you thinking to yourself? | Identifies troubling thoughts |
| 4 | What thought is the most troubling? | Focuses attention on the most troubling thought |
| 5 | What do you feel when you think this? | Reinforces the core CBT principle that thoughts trigger feelings |
| 6 | When you have these feelings, what actions do you take? What do you avoid? | Identifies behaviors that are caused by the cascading effect of thoughts and feelings |
| 7 | Retype the summary of the situation in the following format: Trigger: Thought: Feeling: Behavior: | Synthesizes past reflection by highlighting the connection between the trigger and its manifestations |
| 8 | Do you believe that the initial trigger justifies the intensity of your thoughts and feelings? | Challenges potentially negative thought patterns |
| 9 | How can interpreting the trigger as a challenge rather than a threat alter your response? | Encourages cognitive reinterpretation of the situation |

| 10 | What new perspectives could you adopt to view these challenges as opportunities? | Promotes a reappraisal frame focused on growth |
|---|---|---|
| 11 | How might this change in perspective influence your future reactions to similar challenges? | Encourages transfer of reappraisal strategies to future scenarios |

### *LLM-Driven Implementation*

Motivated by recent advances in LLM-based interventions that support context-aware support [17,22,33], we adapted the sequence in Table 1 to allow content to align more closely with users' specific circumstances. To implement this, we used a prompt engineering approach. The prompt was developed through an iterative process by the same research team. The complete system prompt is provided in Appendix A (see supplementary materials). We structured the prompt using GPT-4o with the following components:

- **Structured reflection process**: The prompt was designed to guide users through a step-by-step conversation aligned with Table 1.
- **Conversational interaction through natural dialogue**: Instead of presenting fixed or survey-like questions, the chatbot was instructed to respond naturally and acknowledge the user's situation. If a response was vague or incomplete, it was instructed to ask clarification questions to gather more detail. The chatbot was also instructed to avoid repeating the same phrases or prompts. The chatbot was designed as a structured facilitator; it did not produce advice, suggestions, or reframings on the user's behalf. The reappraisal questions (Q8-Q11) were designed to prompt users to arrive at these insights through their own reflection, with the chatbot's role being to acknowledge, clarify, and guide rather than to supply content.
- **Reflective instructional prompt**: Prior work has shown that in longer conversations, LLMs may deviate from the intended behavior, even when following the prompt, because they try to adjust to the user's specific responses, clarifications, or context [34]. In our case, this could sometimes lead the model to drift away from the specific question or theme it was meant to address. To handle this, we added an instruction for the model to first generate a response, then check if it aligns with the current question's theme, and revise it if needed [35].

The chatbot was implemented as a text-based conversational agent and was embedded directly within the Qualtrics survey platform. It did not have a named persona or profile picture. The chatbot opened each session with a fixed introductory message welcoming the participant and describing the nature of the interaction (see Appendix A). The language style was conversational and reflective, with a temperature setting of 0.7 to allow for naturalistic variation in responses while maintaining coherence. A screenshot of the interface is provided in Appendix C (see supplementary materials).

### Participants

We recruited information workers from a large US-based technology company with a broad portfolio spanning software, hardware, cloud services, and cybersecurity. Employees based in the USA were invited via email to participate in a voluntary survey. We recruited from this setting because workplace stress is well-documented as particularly prevalent in the technology sector [36,37], and because our affiliation with a technology company provided internal access to a large, geographically distributed employee population.

Participants were informed that the study focused on AI conversational agents for supporting workplace stress scenarios. A total of 100 participants completed the study. They represented multiple gender identities (69 men, 27 women, 4 undisclosed; more options were offered) and varied in age (22 aged 18–35, 61 aged 36–55, 17 aged 56+). Prior to engaging with the intervention, participants completed the 10-item Perceived Stress Scale (PSS; [26]) to assess their stress levels over the past month. The average score was 21.26±3.67, which falls within the moderate stress range (typically defined as 14–26). The PSS served as a baseline measure to characterize participants' general stress levels at the time of the study; no minimum score was required for participation. Scores on the PSS range from 0 to 40, with higher scores indicating greater perceived stress: 0–13 = low, 14–26 = moderate, 27–40 = high.

Participants also reported their prior experience with text-based AI conversational agents: 34% had 6 months to 1 year of experience, 33% had 1 to 2 years, 22% had 2 to 5 years, 6% had more than 5 years, and 5% had less than 6 months, suggesting that most participants had prior familiarity with conversational agents. A detailed breakdown of participant demographics, along with distributions of PSS scores, is available in Appendix B (see supplementary materials).

### Procedure

Participants were invited to engage in an SSI in which they reflected on a workplace stress situation through a conversation with a chatbot. The study was disseminated through a one-time email invitation, and participants could complete the activity at any time of their choosing after receiving it. The intervention was designed as a single, one-time activity to be completed within a continuous sitting; participants were not able to return to complete additional sessions. The chatbot guided them through a series of questions designed to support reflection on their experience. Participants were encouraged to respond genuinely to each prompt. At the end of the interaction, the chatbot indicated that the conversation was complete, after which participants proceeded to the next page. They were informed in advance that the conversation would likely involve around 15 conversational turns, based on the 11 core questions programmed into the chatbot, along with potential follow-up and clarification prompts. Appendix C (see supplementary materials) demonstrates an example conversation between the chatbot and a user. All components of the study, including the chatbot interaction, were completed within the Qualtrics platform.

To assess the impact of the SSI, participants completed a set of self-report questions both before and after the session. To evaluate short-term changes following the intervention, we selected measures that capture both immediate subjective stress responses and theoretically relevant appraisal-related constructs. Specifically, we assessed situational stress intensity, stress mindset, perceived demand, and perceived resources, as these reflect how individuals interpret and respond to stressful situations. All responses were recorded on a 5-point Likert scale, where 1 indicated low agreement or intensity (e.g., ``strongly disagree'', ``very low'') and 5 indicated high agreement or intensity (e.g., ``strongly agree'', ``very high''). The measures included:
**Situational Stress Intensity:** A single question assessing participants' stress level in relation to a workplace situation [8]. We selected this brief measure to capture immediate subjective stress relevant to the focal scenario and to minimize burden within a single-session study.
**Stress Mindset:** Eight statements capturing beliefs about the potential enhancing or debilitating effects of stress (e.g., ``Experiencing this stress facilitates my learning and growth''; ``The effects of this stress are negative and should be avoided'') [38]. These included both positively and negatively framed statements. We included this measure because cognitive reappraisal may shift how individuals interpret the meaning and consequences of stress.
**Perceived Demand:** A rating of how demanding the stressor felt [36,39]. This measure was included as an exploratory outcome grounded in the Job Demands-Resources model [39].
**Perceived Resources:** A rating of the participant's perceived ability to cope with the stressor [36,39]. Alongside perceived demand, this measure captured whether the intervention shifted participants' sense of their own coping capacity [39].

Situational stress intensity served as the primary outcome measure, as it most directly captures the intended effect of the reappraisal activity. The remaining measures -- Stress Mindset, Perceived Demand, and Perceived Resources -- were exploratory, included to provide a broader characterization of participants' stress-related experiences before and after the intervention.

To gather user feedback about the intervention, participants also responded to open-ended questions about how the activity affected their stress levels and their view of the situation, as well as what aspects they liked or disliked. All open-ended questions were provided immediately at the end of the session, within the same Qualtrics survey. These questions included:

- How did this activity affect your stress levels? Feel free to explain in as much depth as you would like.
- Please explain whether and how interacting with the chatbot has changed your view of the stressful situation.

- Please comment on what aspects of the chatbot interaction you liked or found helpful. Why?
- Please comment on what aspects of the chatbot interaction you did not like or did not find helpful. Why?
- Please suggest how we can further improve this activity.

All open-ended questions were mandatory except for the improvement suggestion question, which was optional. All 100 participants provided responses to the four mandatory questions, and 66 provided responses to the optional improvement suggestion question.

Because the study used a one-time, survey-based format, participants who did not complete the study did not submit responses and therefore were not included in the final sample. We were unable to assess dropout rates among participants who may have started but not completed the study.

### Data Analysis

As a feasibility study, our objective was to examine whether engagement with an LLM-based chatbot for cognitive reappraisal could lead to observable improvements in stress-related outcomes, such as situational stress intensity and stress mindset. Early-stage feasibility studies serve a distinct role in intervention research [27]; they help determine whether the conceptual approach is worth pursuing, whether users engage meaningfully with the system, and whether measurable changes occur in theoretically relevant constructs. Demonstrating even modest improvements under these conditions can support the potential viability of LLM-based reappraisal as a foundation for more controlled investigations. To contextualize these quantitative findings, we complement them with qualitative analyses that explore how users experienced and interpreted the chatbot's support, providing insight into the underlying mechanisms and perceived value of the interaction.

We conducted paired Wilcoxon signed-rank tests to assess pre–post differences across the four measures described in the "Procedure" Section. We applied the Benjamini–Hochberg (BH) procedure to account for multiple comparisons.

To complement user-reported scores, we analyzed linguistic features of user–chatbot conversations. Each conversation was divided into three segments (T1 = beginning, T2 = middle, T3 = end), and we examined how predicted stress and sentiment varied across them using three approaches:

**RoBERTa sentiment classifier**: A pre-trained RoBERTa-based model [40] that outputs a probability distribution over *positive* and *negative* sentiment for each conversation segment.
**RoBERTa stress classifier:** A pre-trained RoBERTa-based model [41] that estimates a *stress probability* ranging from 0 to 1 for each conversation segment.
**LLM stress rater:** A prompting-based approach that assigns a stress score to each user message using a structured prompt informed by prior work on LLMs as evaluators [42–46]. The rater outputs a score from 1 to 5, where 1 = no observable stress and 5 = high stress (characterized by pronounced emotional strain or difficulty coping). The rating criteria were developed through a review of prior literature and iterative design, enabling finer distinctions in stress intensity. Further details about the prompt design are provided in Appendix D (see supplementary materials).

While these methods provide useful approximations of users' affective and psychological states, we recognize that their outputs can be shaped by modeling assumptions and may miss contextual nuance. We therefore interpret them as supportive evidence for our quantitative analyses rather than as definitive indicators of users' internal experiences.

For the qualitative analysis, we first anonymized and cleaned participants' responses, then conducted a thematic analysis [47]. The first author independently reviewed all participant responses and developed an initial codebook using an open-coding method. This process reflected an iterative approach that integrated both deductive and inductive techniques: the analysis of RQ1 was guided by prior work on stress reappraisal interventions , while RQ2 was approached inductively, allowing themes to emerge directly from the data. The first author then used axial coding to refine categories and develop higher-order themes. Throughout the analysis, the codes and themes were regularly discussed with other members of the research team to ensure multiple perspectives were considered. The development of themes was reviewed through detailed collaborative discussions. This approach is consistent with qualitative coding practices used in prior studies [48,49].

### Ethical Considerations

This research was reviewed and approved by the Institutional Review Board (IRB) at the technology company where the study was conducted. Given the sensitive nature of the intervention, which included reflection on negative thoughts and potentially distressing life situations, we were attentive to ethical considerations throughout the research process and implemented safeguards to support participant well-being. All participants provided informed consent before beginning the study. To protect confidentiality, they were explicitly requested not to include any personally identifiable information in their responses. Participants were also informed that they could skip any question or withdraw from the study at any time without consequence. Additionally, participants were provided with information about internal support resources in case they experienced stress, anxiety, depression, or emotional discomfort during or after the session.

## Results

Participants wrote an average of 12.81±1.66 messages per conversation, contributed 283.74±243.16 words, and spent 23.09±23.99 minutes engaging with the chatbot. These results suggest that participants were responsive to the prompts and invested a reasonable level of time and effort in composing substantive replies. Building on these observations, we now examine how our study addressed the research questions.

### How does engaging with an LLM-based chatbot for cognitive reappraisal of a stressful scenario impact individuals' situational stress levels?

Participants reported significant reductions in situational stress intensity following the intervention ($\Delta = 0.29 \pm 0.83$, $p = 0.002$, $r_{rb} = 0.54$) and significant improvements in stress mindset ($\Delta = 1.70 \pm 4.37$, $p = 0.002$, $r_{rb} = 0.44$). The latter reflects a shift toward viewing stress as more enhancing than debilitating. Perceived resources showed improvement ($\Delta = 0.17 \pm 0.83$, $p = 0.07$, $r_{rb} = 0.32$), and perceived demand showed a reduction ($\Delta = 0.12 \pm 0.83$, $p = 0.17$, $r_{rb} = 0.22$), although these changes were not statistically significant.

**Table 2.** Pre- and post-intervention means and SDs for all outcome measures. The Improvement column reports change in the beneficial direction, computed as Post − Pre for measures where higher is better (↑) and Pre − Post for measures where lower is better (↓), so that positive values always indicate improvement. Rank-biserial correlations ($r_{rb}$) are likewise signed so that positive values favor the intervention. P-values are corrected using the BH procedure ($\alpha = 0.05$).

| Measure | Pre (Mean ± SD) | Post (Mean ± SD) | Intended Direction | Improvement (Mean ± SD) | W | p-value | $r_{rb}$ |
|---|---|---|---|---|---|---|---|
| **Situational Stress Intensity** | 3.38 ± 0.95 | 3.09 ± 0.91 | ↓ | 0.29 ± 0.83 | 216.0 | 0.002** | 0.54 |
| **Stress Mindset** | 20.71 ± 6.64 | 22.41 ± 7.38 | ↑ | 1.70 ± 4.37 | 994.0 | 0.002** | 0.44 |
| **Perceived Demand** | 3.40 ± 0.91 | 3.28 ± 0.89 | ↓ | 0.12 ± 0.83 | 385.0 | 0.17 | 0.22 |
| **Perceived Resources** | 2.97 ± 0.97 | 3.14 ± 1.00 | ↑ | 0.17 ± 0.83 | 306.5 | 0.07 | 0.32 |

** p < 0.05; ** p < 0.01; *** p < 0.001*

We examined the progression of average negative sentiment and stress scores across conversation tertiles (T1–T3) for all three classifiers, followed by non-parametric statistical analyses. Results are presented in Figure 1 and Table 3. Across all three classifiers, users' messages exhibited a consistent pattern: sentiment became more positive, and indicators of stress decreased from the beginning to the end of the interaction. Friedman tests revealed statistically significant differences across tertiles for each classifier. Pairwise Wilcoxon tests showed significant changes between T1 and T3, as well as between T2 and T3, in all three cases. These findings suggest that as users engaged with the chatbot, their expressions of negative sentiment and stress declined over time, aligning with the observed reduction in self-reported stress scores.

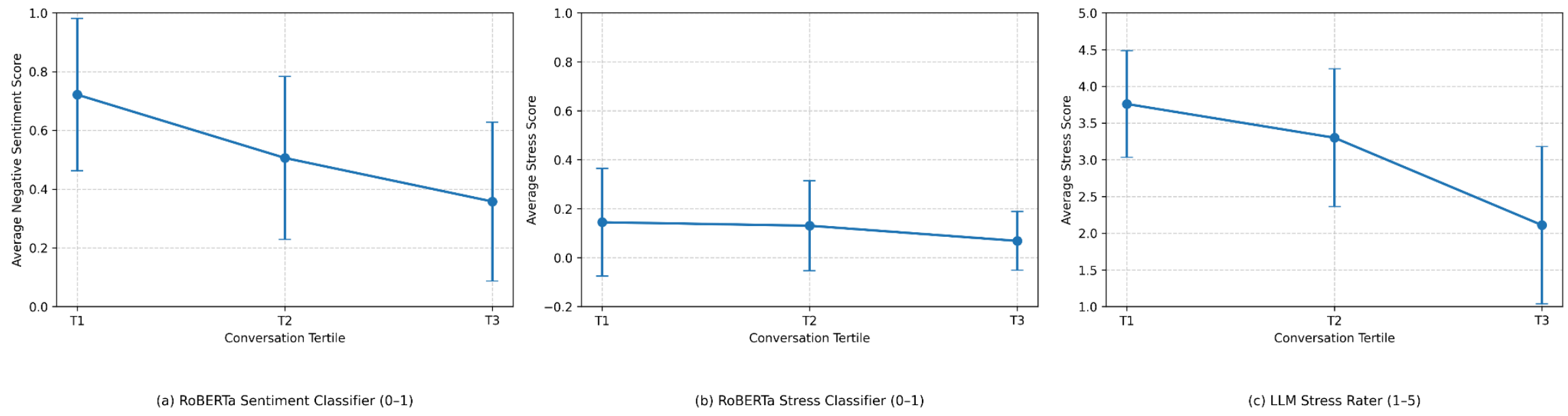

Figure 1: Changes in sentiment and stress scores across conversation tertiles (T1-T3). Error bars represent standard deviations.

| RoBERTa sentiment classifier | | RoBERTa stress classifier | | LLM stress rater | |
|---|---|---|---|---|---|
| Omnibus test: Friedman across T1–T3 | <0.001*** | Omnibus test: Friedman across T1–T3 | <0.001*** | Omnibus test: Friedman across T1–T3 | <0.001*** |
| Pairwise tests (Wilcoxon signed-rank): | | Pairwise tests (Wilcoxon signed-rank): | | Pairwise tests (Wilcoxon signed-rank): | |
| T1 vs T2 | <0.001*** | T1 vs T2 | 0.69 | T1 vs T2 | <0.001*** |
| T1 vs T3 | <0.001*** | T1 vs T3 | <0.001*** | T1 vs T3 | <0.001*** |
| T2 vs T3 | <0.001*** | T2 vs T3 | <0.001*** | T2 vs T3 | <0.001*** |

Table 3: Statistical tests accompanying Figure 1, assessing changes in negative sentiment and stress scores across conversation tertiles (T1 to T3) for all three classifiers. For each classifier, we report an omnibus Friedman test across T1–T3 and three pairwise Wilcoxon signed-rank tests. All p-values are corrected using the BH procedure. Significance levels: * $p < 0.05$; ** $p < 0.01$; *** $p < 0.001$.

### *General Appreciation for Structured Reflection*

Participants in general highlighted the value of the chatbot's step-by-step approach, noting that the ordered prompts helped them organize thoughts that might otherwise remain scattered. P55 described the session as "*a good exercise in laying out the issue, my responses, and getting to why it's that way*", and P42 emphasized how the sequence of questions encouraged structured analysis rather than a quick vent. Several also echoed that the structure offered a clear path through the conversation, making it easier to track what had been covered and where to focus next.

Participants often remarked that the value of the interaction lay in how it helped them arrive at their own insights through deliberate unpacking of the stressor. They acknowledged that the structured opportunity to express themselves brought a sense of relief. The act of "thinking aloud" through writing or guided dialogue was also perceived as familiar and useful. Drawing a comparison to journaling, P15 reflected:

*"This activity did decrease my stress level by allowing for a structured conversation regarding a specific stressor. It helped me reflect on my own thoughts without spiraling into anxiety. Just by explaining and going through the situation it made me feel more calm, similar to journaling but more structured. I have previously liked journal prompts in the past and this feels similar."*

Some participants also appreciated that the structured format helped shift their focus from problems to potential solutions. They noted that the prompts encouraged them to slow down and articulate their thoughts more deliberately. As P89 explained, "*I often struggle for words, but having a chatbot to ask these questions makes me feel like I can take my time to find the right way to say what I want to say.*" These comments suggest that participants valued not just the content of the activity, but the structured way in which it guided their thinking.

### *Reinterpreting Stressful Situations*

The shift observed in stress mindset scores—toward viewing stress as more enhancing than debilitating—was echoed in participants' qualitative comments. Several described how the conversation helped them reassess their situation from a new angle. P79 shared that it "*shifted my perspective just enough to see it more as a challenge than a frustration*", while P21 noted that the chatbot offered an alternative way of thinking when they felt mentally stuck, helping them move beyond unproductive thought patterns.

The conversation logs further reflect this reappraisal process. P4, who began the session feeling overwhelmed and unsupported by colleagues, wrote at Q9 that '*perhaps the anger and anxiety are the result of repeated cycles and maybe I need to consider I am burnt out on the role*' — a reinterpretation that emerged through their own responses to the structured questions. P65, working through anxiety about executing quickly on a difficult technical project, arrived by Q10 at a reframe of the work as '*an exploration, an open mindset that allows me to explore boundaries*', shifting from perceiving the situation as a threat to viewing it as an opportunity for learning.

However, not every participant experienced dramatic shifts in perspective. P30, for example, remarked that "*while it did not change my viewpoint 100%, it was nice to write it out*", suggesting that even partial reframing can be helpful. Others, including

P15 and P48, described subtle changes in how they interpreted or emotionally responded to the stressor. These comments highlight the potential of the interaction to support gradual cognitive shifts, helping participants reinterpret their stressors and relate to stress in a more constructive way. It is worth noting that these reframings were generated by participants themselves through the structured reappraisal questions; the chatbot's role was to guide and facilitate this process rather than to supply alternative interpretations directly.

### *Identifying Concrete Next Steps*

Beyond perspective shifts, some participants described how the dialogue helped translate reflection into concrete action. P66 appreciated "*the structured approach to break down the root of my stress and ideate on solutions*", while P83 noted that the activity helped them "*channel stress into action*." P70 shared that although their stress level did not change, the activity "*helped to structure my approach to the situation*", and P76 commented that the conversation "*helped lead me to some useful options to help me fix my problem*." These comments suggest that, for some participants, the value of the activity lay not only in processing emotions and perspectives but in arriving at a clearer sense of what to do next.

The conversation logs further illustrate how this process unfolded during the later stages of the activity. Participants who initially described interpersonal tension or coordination difficulties sometimes progressed toward specific action plans by the final reappraisal prompts. For example, P73, who began the session feeling overwhelmed by disorganized planning and strained colleague relationships, later described intentions to foster better communication and establish more regular check-ins with their team before the next planning cycle. Similarly, P61, who entered the conversation frustrated by a difficult team dynamic, moved toward articulating concrete steps such as involving a new team member in daily discussions and investing in a stronger working relationship. P59, who entered the session anxious about presenting to a large group of external customers in a non-native language and worried about losing the audience's attention, arrived by the final prompts at a concrete plan to use auto-captioning technology in future presentations and to build in pauses for audience questions. These examples suggest that the structured questioning sequence could help some users move from diffuse stress toward more actionable coping plans.

### *Validation of Personal Experiences*

Some participants described the chatbot as helpful in recognizing or affirming their existing efforts and emotional responses. Even when the chatbot did not introduce new ideas, it was still seen as useful for reinforcing what participants were already feeling or doing. As P16 explained, "*Just because it didn't change my view or give me any new data to think about doesn't mean it wasn't useful… it's validation that there isn't something I'm missing, that I've explored every opportunity and tried everything I can.*"

Others similarly appreciated moments where the chatbot reflected back their current state of mind. P74 noted, "*It was great at validating my feelings and guiding me in its approach, even when I was reluctant to think about a solution.*" P82 shared that "*the chatbot acknowledged my efforts as productive*" in dealing with a difficult situation, and P92 mentioned it "*affirmed steps planned.*"

For a few, this sense of reinforcement was seen as limited. As P46 commented, "*Lots of validation. It might have been better if it had challenged me.*" Still, for many participants, the interaction was valued for offering a sense of confirmation rather than redirection.

### *Appreciation for Non-Judgmental Interaction*

Several participants appreciated the chatbot's non-judgmental tone, describing it as a space where they could express themselves freely. For some, this quality made the interaction feel less pressuring than conversations with people. P42 remarked that it "*allowed me to vent, without fear of judgment*", while P65 shared, "*I liked that it was objective.*"

Some others described the experience as supportive precisely because it did not involve being evaluated. P66 emphasized the benefit of having "*an unbiased conversation" that helped them "focus on a task at hand.*" This perception of the chatbot as non-judgmental contributed to some participants' willingness to open up or think through the situation more deliberately.

### *Perceived Misalignment with Context or Needs*

Some participants described instances when the chatbot's responses felt disconnected from the realities of their situation. In particular, a few noted that the interaction seemed to assume ideal conditions, such as collaboration or personal control, that did not reflect their experiences. As P36 explained:

"*The chatbot assumes that all the people involved in the situation are reasonable, collaborative, and want to reach a solution that's acceptable to everyone. But the nature of my situation is that the other team has already established that they are none of these things.*"

P94 similarly expressed that the prompts did not adequately reflect the complexity of their stressor, stating, "*I felt that I needed to double down and explain more about the stressful situation, to say it's not that easy.*"

Some participants also described discomfort with having to revisit a stressful situation in detail, particularly when they had already discussed it elsewhere. P59 remarked, "*I already had to explain the situation to myself, then my team, and now this bot wants to know about it too, causing me to relive the situation in my mind.*" Others felt that, despite going through the interaction, there was no tangible outcome. As P24 noted, "*I dislike the fact that it goes nowhere and there is no action to be done.*"

These reflections suggest that for some participants, retelling the situation without any follow-up or resolution made the experience less satisfying.

## What design tensions arise when individuals interact with an LLM-based chatbot to cognitively reappraise a stressful scenario?

### *Scripted Structure vs. Conversational Naturalness*

As discussed in the previous sections, participants noted that the structured nature of the chatbot's prompts helped them consider alternative perspectives and identify potential solutions. However, several participants also reflected on how this structure may have come at the expense of a more natural, conversational flow. P6 remarked that "*it wanted to ask certain questions as opposed to engage in conversation*", highlighting a perception that the interaction was being guided more by a fixed sequence than by mutual exchange.

Some participants further sensed that the chatbot was following a script, which contributed to a feeling of artificiality. As P75 put it, the interaction felt "*not human*", due to what appeared to be a scripted, predetermined flow. While certain elements of the structure, such as follow-up questions or requests for clarification, were appreciated by some, they were not universally welcomed. For example, P94 noted that these follow-ups "*helped pull out some important information*", but others, like P96, found them repetitive and frustrating: "*I didn't necessarily like that it always yielded additional questions, some of which I felt I had already answered earlier in the conversation*." Together, these responses underscore a tension between the benefits of a structured, goal-directed dialogue and the desire for a more natural and responsive conversational experience.

### *Balancing Contextual Understanding and Conversation Length*

Some participants expressed a desire for more questions to allow them to elaborate on their situation—particularly when their stressors were complex or nuanced. P7 noted the difficulty of condensing their experience, remarking, "*It's the writing it down, trying to cover the important topics with brevity. Tough nut to crack.*" Similarly, P57 wished for a longer "*getting to know each other*" period to help the chatbot better personalize its responses.

Conversely, some participants felt that even the current level of interaction was somewhat extended. P12 commented that the exchange lasted "*much longer than I cared for*." They expressed that the number of questions, often accompanied by clarifications and follow-ups, made the experience feel more effortful than necessary. This concern was particularly salient among participants dealing with less intense stressors, where a longer interaction felt disproportionate. As P23 explained, "*My stress trigger was basic and didn't need to have that many turns*." These perspectives reveal a core design tradeoff: while deeper context can enable more tailored and relevant responses, excessive questioning may lead to user fatigue or disengagement in some cases.

### *Perceptions toward AI and Artificial Empathy*

Although the study did not explicitly ask about participants' attitudes toward AI, some participants shared comments that reflected preexisting skepticism about the use of AI and technology for well-being support. Several participants indicated that their discomfort stemmed not from the content of the conversation, but from their underlying beliefs about what makes an interaction feel helpful. P29 shared that "*the fact that it was a chatbot made it annoying*", and P37 expressed discomfort with "*the fact that it was not a real person*." These comments suggest a reluctance to view non-human agents as appropriate partners for discussing personal or emotional matters. P51 further expressed mistrust toward the system, stating that they do not trust a software entity that cannot feel emotions.

These views extended to how participants interpreted the chatbot's attempts at empathy. Some reacted negatively to what they perceived as artificial emotional cues. P14 stated, "*The whole mimicking empathy is annoying*", and P41 described the interaction as "*trying too hard, like mimicking empathy but not quite getting there*." In contrast, others responded more positively to the chatbot's empathic tone. P44 noted that the message felt "*kind*", even though they knew it was automated. P58 similarly shared, "*I could tell it was a bot, but it still felt like it was trying to be understanding.*" These contrasting reactions highlight how users' preconceptions about AI—and particularly their expectations around empathy from AI—impacted their perceptions of AI support.

## Discussion

In this work, we present a design for a cognitive reappraisal intervention that leverages LLMs to provide conversational interaction. Our SSI incorporated a structured conversational scaffold to guide the reappraisal process, enabling individuals to revisit initial interpretations, examine different aspects of their challenges, and consider alternative explanations. This combination of a structured SSI format with LLM-driven conversational flexibility addresses a well-documented limitation of scripted tools, which typically cannot adapt to the varied ways users describe their experiences. Our deployment showed that this approach can retain many benefits of past cognitive reappraisal interventions, including reducing negative emotions, increasing clarity about one's situation, and encouraging reflection on potential growth opportunities [2,8,22]. However, integrating LLMs introduced several design tensions, particularly in balancing contextualization with the number of questions asked, maintaining a structured flow while preserving conversational quality, and managing perceptions of AI tone and empathy.

In the remainder of this section, we reflect on the key findings of our study, discuss how they align with or extend existing literature, and propose relevant directions for future research. We conclude with a discussion of the limitations of the study.

### Principal Results

Our findings suggest that the LLM-based cognitive reappraisal activity can retain many of the benefits associated with more traditional reappraisal techniques.

Participants reported reductions in situational stress and improvements in stress mindset after engaging with the intervention. These shifts were supported by the structured nature of the activity, which guided users in breaking down the elements of their stressful experiences. This process enabled them to reassess their initial interpretations and consider alternative perspectives—an outcome consistent with prior research on the role of cognitive reappraisal in fostering more adaptive emotional responses [1,2,6,50].

In addition to promoting perspective shifts, the activity helped some participants reflect on possible actions they could take to address their situations. Several described recognizing specific next steps or reframing their experiences as opportunities for growth [2,22]. These qualitative observations were further corroborated by three independent classifier-based measures, which showed consistent reductions in expressed stress and negative sentiment across conversation tertiles, providing convergent evidence beyond self-report. This highlights how structured LLM-enhanced interventions can support cognitive reframing while helping users identify concrete ways to respond to their stressors.

Even when substantial cognitive shifts did not occur, participants described experiencing subtle emotional changes through their engagement with the structured, LLM-enhanced activity. In particular, the process of *externalizing* a stressful situation (i.e., expressing internal thoughts and emotions outwardly [51]) was seen as helpful in itself. Several participants likened the experience to journaling, which can offer a sense of release and promote clarity [52–54]. These findings suggest that the value of reappraisal-based interventions may extend beyond cognitive reframing, by also offering structured opportunities for reflection and emotional processing through expressive dialogue.

At the same time, participants' responses surfaced a tension between the benefits of a structured sequence and the desire for a more naturalistic conversational flow. While the structure was widely valued for organizing reflection, some described the interaction as feeling scripted, particularly when follow-up questions appeared to repeat information already provided. Designing LLM interactions that move smoothly between structured progression and responsive acknowledgment, without drifting from the intended sequence, should be an active research area for future work.

However, we also observed that cognitive reappraisal may not be appropriate in all scenarios. Participants pointed out that, although support was directed at a single individual, the source of stress could be more complex and involve multiple people within their network. This challenge is not unique to cognitive reappraisal but reflects a broader limitation in much of the literature on health interventions, which often emphasizes individual-level support [55,56]. In response to this gap, a growing line of research is exploring group-level interventions that engage multiple individuals in jointly addressing shared or relational challenges [57–59]. Building on this momentum, future work could explore the design of technology-mediated cognitive reappraisal interventions that support group-level reflection [60]. For example, digital platforms could facilitate shared appraisal by prompting teams to surface collective stressors, reflect on interpersonal dynamics, and co-construct alternative interpretations. By embedding reappraisal within team interactions, these technologies may have the potential to foster shared understanding, strengthen social support, and address the relational nature of stress more effectively than individually oriented interventions.

### Considerations for Broader Applications

Our designed activity can function as an SSI, offering a brief, structured opportunity for individuals to engage in cognitive reappraisal in response to everyday stressors. SSIs have been shown to produce meaningful psychological benefits with minimal time commitment [24,25], so they could be well-suited for integration into demanding professional environments. We emphasize that LLM-based cognitive reappraisal interventions need not follow the exact structure used in our study. Our results point to the tradeoff between contextual depth and conversation length. Some participants wished for more opportunities to elaborate on complex stressors, while others found the number of prompts excessive when their stressor felt straightforward. Future LLM-based interventions could explore adaptive mechanisms that modulate the depth and length of the interaction based on the complexity of the stressor [9], while ensuring that shortening the interaction does not come at the cost of the reappraisal process itself.

We note that while this study was conducted in a workplace setting, the cognitive reappraisal sequence was designed to be applicable across a broad range of stressful situations rather than tailored exclusively to occupational stressors. The workplace context shaped our recruitment, participant population, and the types of stressors participants brought to the activity, but the generality of the intervention was intentional: cognitive reappraisal as a strategy is not inherently context-specific, and a broadly applicable sequence may be more easily adapted across deployment settings in future work.

We also note that these interventions alone are not sufficient to address the broader challenges of well-being and may serve as modular components within a more comprehensive framework alongside other strategies [29]. These SSIs could be embedded within larger DMH programs spanning multiple weeks and complement practices such as mindfulness activities or psychoeducational content [29]. When our activity is repeated over time, future work should explore how to maintain a sense of freshness and avoid monotony by incorporating variations in structure, tone, or content of the responses. However, the use of LLM-generated content for more severe mental health conditions, including post-traumatic stress disorder or suicidal ideation, would likely require additional safeguards such as clinical review and supervision prior to delivery.

A related design concern that warrants attention as these interventions scale is the tendency of LLM-based systems toward over-validation [61,62]. While participants in our study generally appreciated the non-judgmental and affirming tone of the chatbot, some noted that the interaction offered little challenge or pushback. This reflects a broader pattern documented in the literature, where LLMs acting in therapeutic or supportive roles have been found to over-validate users' beliefs without promoting the kind of cognitive flexibility that can underlie reappraisal [62]. Future design efforts should explore how to calibrate the balance between validation and constructive challenge, for instance by incorporating adaptive feedback mechanisms that respond to the depth and quality of a user's reappraisal attempts.

More broadly, as LLM-based mental health tools become more widely adopted, important questions arise about their relationship to professional care. These tools have the potential to extend support to individuals who would not otherwise access services, but if users come to rely on AI for emotional support in lieu of professional help, there is a risk of eroding motivation to seek human care precisely among those who may need it most. Future work should examine how LLM-based interventions can be designed to complement rather than substitute professional relationships, for example by incorporating referral pathways or escalation mechanisms for users who may require more intensive support [19].

### Perceptions of AI Tone and Empathy

Participants shared a range of preferences and reactions regarding the tone of the chatbot and its attempts at conveying empathy. Many appreciated the non-judgmental tone, noting that it offered a space where they could articulate their thoughts without fear of being judged, similar to prior benefits of non-judgmental care [63,64]. At the same time, the structured sequence of prompts, while helpful for reflection, sometimes made the interaction feel more like a scripted exchange than a fluid conversation. This sort of perception can reduce the potential of DMH interventions [65].

A related tension emerged around the expression of empathy. While prior work has emphasized the importance of empathetic responses in mental health interventions [33,66], some participants were skeptical of AI's capacity to provide genuine emotional support. In some cases, these preexisting beliefs contributed to how participants interpreted the chatbot's attempts at empathy, with some finding them helpful and others perceiving them as forced or artificial.

Future work could address these tensions by reducing the perceived scriptedness of the interaction and more deliberately shaping the agent's persona. A growing body of research has explored the use of human-like traits in conversational agents, and incorporating elements such as tone, conversational style, and empathy level based on models of anthropomorphism [67] or personality frameworks (e.g., the Big Five [68]) could improve user satisfaction. These traits could be either predefined or adapt dynamically based on user preferences and expectations [69]. Designing agents that are also sensitive to how users perceive AI's role in providing support, particularly in sensitive contexts, may help ensure that the tone and level of empathy feel appropriate rather than artificial.

## Limitations

Our study has several limitations. First, although pre–post measures allowed us to capture within-subject changes, such designs may be susceptible to response shift bias [70] or demand characteristics [71]. Participants' awareness of being measured before and after the intervention may have influenced how they interpret or report their experiences. In addition, several outcomes were assessed using brief or single-item self-report measures selected to reduce participant burden in a single-session study, which may provide less precision than longer validated instruments. Future work deploying similar interventions in a longitudinal or multi-session format should prospectively track engagement and assess reasons for non-completion. Additionally, as a feasibility study, the absence of a control condition means we cannot isolate the contribution of the structured reappraisal sequence from the general effects of expressive writing or conversational engagement.

In our study, we used the GPT-4o model, a state-of-the-art LLM at the time of the study, to generate responses for the reappraisal activity. We designed custom prompts to structure the interaction, which participants generally found appropriate and helpful. However, different models may vary in how they respond to similar prompts, and these differences could influence the perceived quality and usefulness of the interaction. Smaller or less capable models may also struggle to interpret user input or maintain coherence across multi-turn conversations. As such, caution is warranted when generalizing the effectiveness of our prompt design or findings across other LLMs. Future research should examine how different models perform in similar interventions and explore alternative strategies, such as fine-tuning, to ensure consistency and relevance in generated content.

Our sample was drawn from a single large technology company, which may limit the generalizability of our findings to other populations and organizational contexts. Future work should examine whether similar results hold across different industries, demographic groups, and cultural settings, and should also prospectively collect timezone and contextual information to examine how the time and setting of engagement influence outcomes. Additionally, future longitudinal deployments should assess reasons for non-completion to better understand barriers to sustained use.

Finally, our study primarily examined participants' perceptions of the LLM-enhanced cognitive reappraisal activity—their reflections on the experience and its perceived usefulness. We did not systematically evaluate the quality or fidelity of the chatbot's responses. While the activity was designed to follow a structure informed by evidence-based psychological strategies, future work should assess whether the generated content aligns with those strategies and avoids content that may be inappropriate or counterproductive. Involving clinicians or domain experts in reviewing responses could help ensure that the content remains consistent with established psychology principles.

## Conclusions

This study demonstrates the promise of LLMs in delivering structured, adaptive support for stress. Our single-session cognitive reappraisal intervention helped employees reinterpret stressful experiences, leading to significant reductions in immediate stress and improvements in stress mindset. Participants appreciated the guided structure, which offered clarity and validation, while also enabling personalized reflection through contextual responses by the chatbot.

At the same time, integrating LLMs surfaced important design tensions—particularly around the balance between contextualization and user effort, and between structured guidance and conversational naturalness. These insights underscore the need for careful orchestration of structure and flexibility in LLM-based mental health tools. Future work should explore how such interventions can evolve beyond single sessions, adapt dynamically to individual and organizational contexts, and extend to other domains. Our findings contribute both empirical evidence and practical guidance for designing LLM-enhanced DMH interventions.

## Acknowledgements

We acknowledge Kori Inkpen, Andy Wilson, Kael Rowan, Ann Paradiso, and Jieun Kim for their help in testing early versions of the research prototype. Ananya Bhattacharjee would also like to thank Joseph Jay Williams for motivating him to pursue research on reflection and mental health.

## Conflicts of Interest

None declared.

## Abbreviations

JMIR: Journal of Medical Internet Research
DMH: Digital Mental Health
HCI: Human-Computer Interaction
LLM: Large Language Model
SSI: Single Session Intervention

## Multimedia Appendices

URL: http://asset.jmir.pub/assets/d693529098300f5b97216c7d8ae3ef19.docx